\documentclass[a4paper,twocolumn,superscriptaddress,prl,floatfix]{revtex4-1}
\usepackage{epsfig}
\usepackage{epstopdf}
\usepackage{graphicx}
\usepackage{amsfonts,amsmath,latexsym,amssymb}

\begin{document}

\newcommand{\DyO}{Dy$_2$Ti$_2$O$_7$}
\newcommand{\HoO}{Ho$_2$Ti$_2$O$_7$}

\title{Charge ordering in a pure spin model: dipolar spin-ice}

\author{R. A. Borzi}

\affiliation{Instituto de Investigaciones Fisicoqu\'\i{}micas Te\'oricas y Aplicadas UNLP-CONICET and Departamento de F\'\i{}sica,
Facultad de Ciencias Exactas, Universidad Nacional de La Plata, 1900 La Plata, Argentina}

\author{D. Slobinsky}

\affiliation{Instituto de F\'{\i}sica de L\'{\i}quidos y Sistemas Biol\'ogicos, UNLP-CONICET, La Plata 1900, Argentina}

\author{S. A. Grigera}

\affiliation{Instituto de F\'{\i}sica de L\'{\i}quidos y Sistemas Biol\'ogicos, UNLP-CONICET, La Plata 1900, Argentina}
\affiliation{School of Physics and Astronomy, University of St.\  Andrews, St Andrews KY16\ 9SS, United Kingdom}

\date{\today}

\begin{abstract}
We study the dipolar spin-ice model at fixed density of single
excitations, $\rho$, using a Monte Carlo algorithm where processes
of creation and annihilation of such excitations are banned.   In
the limit of $\rho$ going to zero, this model coincides with the
usual dipolar spin-ice model at low temperatures, with the
additional advantage that a negligible number of monopoles allows
for equilibration even at the lowest temperatures. Thus, the
transition to the ordered fundamental state found by Melko {\emph et
al} in 2001 is reached using simple local spin flip dynamics.  As
the density is increased, the monopolar nature of the excitations
becomes apparent: the system shows a rich $\rho$ vs. $T$ phase
diagram with ``charge" ordering transitions analogous to that
observed for Coulomb charges in lattices. A further layer of
complexity is revealed by the existence of order both within the
charges and their associated vacuum, which can only be described in
terms of spins --the true microscopic degrees of freedom of the
system.
\end{abstract}
\maketitle

Defects are the entropic antagonists of the perfectly ordered state,
while at the same time they are inextricably linked to it
\cite{chaikin2000principles}. Topological defects are particularly
stable forms of disorder which, when dense, can lead to higher
hierarchies of order. Thus, for example, the interaction between
vortices in a superconductor can form an Abrikosov lattice or other
forms of vortex matter \cite{Blatter1994}; these phases, in turn,
will have their own topological defects. The
Berezinskii-Kosterlitz-Thouless transition (BKT), dealing with the
unbinding of vortex-antivortex pairs
\cite{Berezinskii1971,Kosterlitz1973} is another example. Among the
endless variety of these defects \cite{Mermin1979}, a new kind of
fractional point-like topological excitation --magnetic charges or
\emph{monopoles}-- was proposed theoretically \cite{Castelnovo2008}
and tested experimentally \cite{Morris2009} in the spin-ice
compounds.   Given their analogy with electrical charges
\cite{Castelnovo2008,Giblin2011}, a rich behaviour is to be expected
at low temperature ($T$) for high monopole \emph{number density}
($\rho$). These are two conditions that are very difficult to
achieve simultaneously in spin-ice, and require a very fine tuning of the
parameters of the Hamiltonian.  In this work we take the alternative
path of studying the full $\rho$ -- $T$  phase diagram in a spin ice
system by  externally fixing the density of magnetic monopoles.

Controlling the density of topological defects is a clean way of
highlighting their essential role in determining some ordered
phases. During the eighties and nineties, for example, the
importance of vortex strings in the 3D XY-model and of ``hedgehog''
point defects in the Heisenberg model phase transition was clearly
shown using this strategy \cite{Kohring1986,Lau1989}.  We take a
similar approach in this letter: we fix the number of defects, but
keep the model unbiased and simulate the spin-ice system not in
terms of the effective degrees of freedom (the monopoles), but in
terms of the individual spins. The strength and beauty of the
monopolar picture of this magnetic system appears reinforced by our
finding of two phases, which can be understood in terms of the
different types of ordering of the attracting monopoles
(i.e. \emph{charge-like} degrees of freedom). Adding to this
remarkable result, our perspective shows in a unified view the
presence of more subtle forms of order: they are related to the many
different ways in which both the monopole-free system (the
\emph{monopole vacuum}) and a perfect crystal of single monopoles
can be assembled in terms of their constituent spins. These spin
degrees of freedom that are not taken into account in the monopolar
picture can be thought in this context as \emph{internal} degrees
of freedom of the monopolar charges and vacuum.  The close
relationship between these findings and previous results in spin ice
is discussed.

The magnetic properties of spin-ice (SI) materials at low
temperatures are well described by the dipolar model
\cite{Bramwell2001a,Melko2004}, in which nearest neighbours
exchange, $J$, and long ranged dipolar interactions with coupling
constant $D$ --both measured in Kelvin-- are taken into account in
the Hamiltonian

\begin{multline}
\frac{\mathcal{H}}{T}=\frac{D}{T} \biggl(\frac{J}{3D} \sum_{<ij>}
S_i S_j + \\ + a^3 \sum_{(i,j)} \left[ \frac{\hat{e}_i \cdot
\hat{e}_j}{|{\bf{r}}_{ij}|^3}
           - \frac{3(\hat{e}_i \cdot {\bf{r}}_{ij}) (\hat{e}_j \cdot {\bf{r}}_{ij}) }{|{\bf{r}}_{ij}|^5} \right] S_i S_j \biggl).
\label{dipolar}
\end{multline}

Here, the magnetic moments (${\boldsymbol \mu}_i$) occupy the sites $i$ of a
pyrochlore lattice, separated by distances $|{\bf{r}}_{ij}|$. They
reside in the vertices of corner-shared tetrahedra (see
Fig.\ 1) of side $a$  and behave as Ising-like spins
(${\boldsymbol \mu}_i = \mu S_i \hat{e}_i$, with $S_i=\pm1$), constrained
to point along the $\langle 111 \rangle$ directions $\hat{e}_i$. 
When the effective nearest neighbours interaction 
$J_{\rm eff} = J/3 + 5/3 D > 0$, $D = \mu_0 \mu^2/4 \pi a^3$,
the \emph{spin-ice rule} is enforced: two
spins should point in and two out of a tetrahedron to minimise its
energy. This rule, combined with the lattice geometry makes SI a
magnetic analogue of water ice, with a similar residual entropy
\cite{ Bramwell2001a}. A violation of the local law implies the
creation of a defect, or {\em monopole}, sitting in the tetrahedron
with a magnetic charge proportional to the divergence of the spin
vectors \cite{Castelnovo2008}. The number of defects at fixed
temperature is thus regulated by the magnitude of $J_{\rm eff}/D$.
In the currently known SI materials, it leads to moderately
correlated monopole fluids \cite{Zhou2011,Zhou2012}. Material design
or the application of external pressure can be used to strengthen
the correlations, revealing new aspects on these systems that we set
out to determine by numerical simulations.

Here we have used the Monte Carlo technique to simulate the dipolar SI
model (Eq.\ 1) with Ewald summations. We modified the dynamics so that
we can have full and independent control over the temperature $T$, which
we measure in units of $D$, and over the density of magnetic charges,
$\rho$. Starting from a random configuration perfectly satisfying the
ice rules, we first flip enough spins to reach the desired number
of positive and negative \emph{single} excitations per tetrahedra
without allowing any double charge excitation.  We then
follow the usual single flip Metropolis algorithm with an additional
constraint: we forbid spin flips which either create or destroy single
defects, preserving detailed balance. In other words, we work in a
statistical ensemble with constant number of single monopole defects,
instead of fixed chemical potential
\cite{Jaubert2009}.  Other details of the simulations can be
found in the supplementary information section (Sup.\ Info.).

The curves in Fig.\ 1 represent our results for the molar specific heat
$C_V$ as a function of temperature at low $\rho$ and linear lattice
size $L = 3~{\rm unit~cells}$. The
limit $\rho \ll 1$ is particularly important, since at very low $T$
we expect our monopole conserving model to coincide with the usual
dipolar SI model. $\rho = 0$ is taken as the minimum non-trivial
number of conserved monopoles (two, of opposite sign). At this
concentration, the evolution of the system consists in the
exploration of the states belonging to an almost perfect SI manifold
(the exponentially degenerate set of two in / two out states) by
means of the random wanderings of the two defects. Although no phase
transition is expected in terms of charges, the specific heat in
Fig.\ 1 shows a sharp peak for $\rho = 0$, centred at $T_V \approx
0.13 D$. An identical feature in $C_V$  was found by R. Melko and
collaborators in the usual SI dipolar model using a multiple spin
flip ``loop'' algorithm \cite{Melko2001a}. They identified the peak
at $T_V$ with a first order transition to a SI ordered ground state,
and proposed an order parameter ($\Psi$) to account for this order
\cite{Melko2001a}. In the present context, this ordering in the spin
system with a virtual absence of monopoles should be interpreted as
a change of the internal state of the {\emph vacuum of magnetic
charges}. The lower inset to Fig.\ 1 shows how $\Psi$ grows below
$T_V$ in our simulations. The quantitative agreement
\cite{Melko2004} --which holds also for other quantities, including
the energy corrected for the presence of a pair of monopoles--
confirms that this minimum number of conserved defects is sufficient
to allow the dipolar model to equilibrate, even when the evolution
is simulated through a simple single spin-flip algorithm.  This
result suggests that the extreme paucity in monopole excitations at low
temperatures is a major factor in the spin freezing observed experimentally.

\begin{figure}[h]
\centerline{\includegraphics[angle=0,width=\columnwidth]{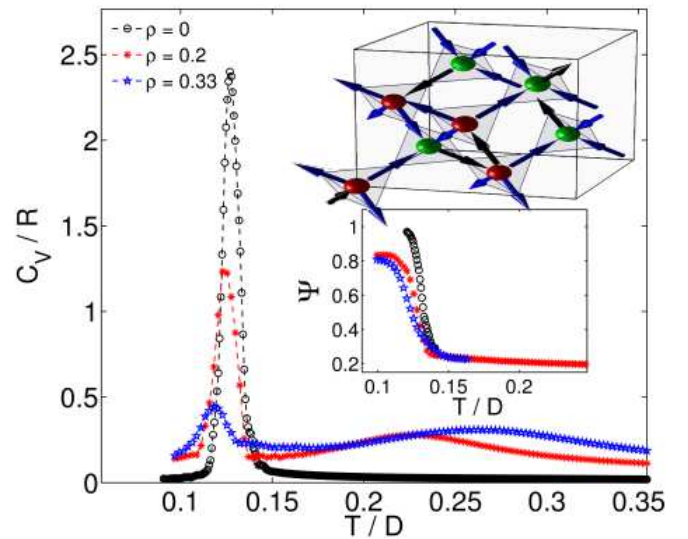}}
\caption{Molar specific heat $C_V$ as a function of $T$ for three
different monopole densities ($\rho$).  The peak in $C_V$ at $T_V$
marks the first order phase transition to an ordered spin phase. The
condensation of the monopoles into a magnetic crystal is signaled by
a broader peak at a higher temperature $T_C$. \emph{Top inset.}
Cubic unit cell for spin-ice materials. The Ising-type spins (shown
as black arrows for \emph{in} and blue for \emph{out} of ``up''
tetrahedra) illustrate one of the ground state configurations found
for the monopole crystal. Both type of single monopoles are plotted
as coloured balls, which conform a ``ionic'' crystal. \emph{Lower
inset.} Temperature dependence of the order parameter for the
spin-ice ground state \cite{Melko2001a}.} \label{fig1}
\end{figure}

An inspection of Fig.\ 1 and its lower inset shows that --albeit to
a smaller extent, and affected by difficulties in equilibrating and
finite size effects-- the vacuum of charges also orders below $T_V$
for non-negligible $\rho$.  More interestingly, these concentrations
show a second (wider) maximum in $C_V$, occurring at a temperature
$T_C(\rho)$ higher than $T_V$. Both $T_C(\rho)$ and the height of
the peak tend to increase with $\rho$, hinting a connection with the
onset of ordering of the magnetic charges. We expect the Coulomb
attraction to favour the clustering of monopoles in the zincblende
structure, where one type of monopole has a greater tendency to
occupy either the up or down tretahedra sublattices. In analogy with
the staggered magnetisation for antiferromagnetism, we use the
\emph{staggered charge density}, $|\rho_S(T)|$, to quantify this
type of alternating or \emph{staggered charge order} (SCO). We define
$|\rho_S(T)|$ as the average of the modulus of the total magnetic
charge in up tetrahedra per sublattice site per unit charge. Figure
2 shows $|\rho_S(T)|$ as a function of temperature for different
sizes $L$ and fixed density $\rho = 0.1$. We can see that
$|\rho_S(T)|$ tends to $\rho$ at low temperatures, decreasing to a
small but finite value at high temperatures that scales as $L^{-3/2}$
(as expected for the average of the modulus of a random variable of
zero mean value, see Sup.\ Info.).  The transition becomes
sharper with $L$, while the temperature of its steepest slope
increases. The top inset to Fig.\ 2 shows the fluctuations in energy
($C_V$) and in $|\rho_S(T)|$ ($\chi_S$) measured at the temperature
$T_C(L)$ at which they peak. Both tend to increase proportional to
the volume of the system, while $T_C(L)$ vs. $1/L$ displays a linear
behaviour (lower inset), indicating a first-order phase transition
\cite{Binderuber}. We have also measured histograms for the
\emph{local} monopole number density, $\rho_{loc}$, which become
bimodal below $T_C$, with peaks near $\rho_{loc} = 1$ and
$\rho_{loc} = 0$. This signals that in addition to developing a staggered
ordering with a net charge
in each sublattice, the system phase separates into a dense
arrangement of monopoles (a \emph{ionic crystal of magnetic
charges}) and a ``fluid'' phase with a very low local concentration of
monopoles.

\begin{figure}[h]
\centerline{\includegraphics[angle=270,width=\columnwidth]{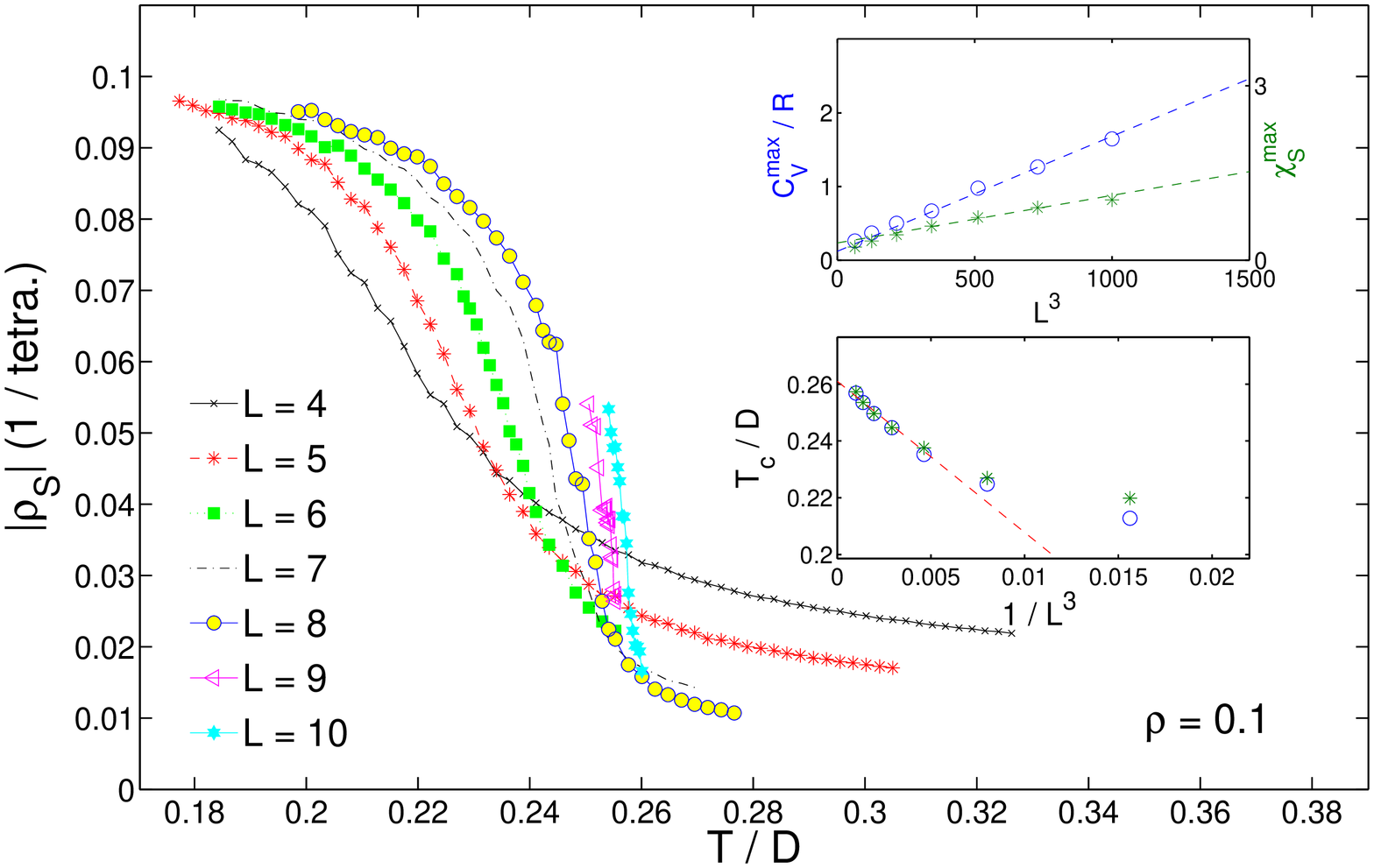}}
\caption{Average modulus of the staggered density $|\rho_S|$ as a
function of $T$, for $\rho = 0.1$. We focus near the crystallisation
transition. On increasing the system size $L$, the order parameter
decreases more abruptly near the transition temperature $T_C(L)$.
The fluctuations in energy (blue trace with open circles) and those
in $|\rho_S|$ (green asterisks) increase with $L^{3}$ (\emph{top
inset}) and $T_C$ evolves with $L^{-3}$ (\emph{lower inset}), as in
a first order transition. } \label{fig2}
\end{figure}

The top part of Fig.\ 3 shows $C_V$ as a function of temperature for
a wide range of $\rho$, and $L = 4$. A small peak is still
noticeable at $T_V$, even for high $\rho$. The height of the cusp at
$T_C$ grows with $\rho$ for small concentrations, as a consequence
of the increase in the relative fraction of crystalline phase being
formed. At $\rho = 0.3$ the cusp becomes wider, and eventually
resolves into two peaks for $\rho \geq 0.5$.  This peak bifurcation
corresponds to the decoupling between the onset of the staggered order
and monopole crystallisation. The low temperature peak is linked
with the second phenomenon, and has a very modest evolution with
$\rho$, occurring always at a temperature (which we call $T_C$)
below $0.32 D$. The opposite is true for the second cusp, that peaks
at temperature $T_S$ going above $0.7 D$ for $\rho > 0.6$. To show
this explicitly, we plot $\rho_S(T)$ in the bottom part of Fig.\ 3,
for the same values of $\rho$ as the upper panel. We observe that
while for small $\rho$ the temperature at which $\rho_S$ is steepest
correlates with $T_C$, it clearly follows the behaviour of $T_S$ for
$\rho > 0.3$. In other words, the position of the high temperature
peak, $T_S$, marks the transition to a phase, with long-range 
staggered charge-ordering but no phase coexistence. It is only 
below $T_C$ that a magnetic crystal separates from a very low vapour
pressure fluid. Between $T_S$ and $T_C$ the average local density is
homogeneous, fluid-like, but with a finite tendency $\rho_S$ for
positive and negative charges to occupy preferentially separate
sublattices.

We summarize most of the results found on this work in Fig.\ 4. It
shows, projected into the $\rho$ vs. $T$ plane, the phase diagram
for our model, drawn on top of an interpolated contour plot of the
specific heat data. The low temperature dome we have labeled as
$T_C(\rho)$ corresponds to the first order transition below which a
monopole crystal with staggered order coexists with a low
vapour-pressure gas \cite{Note1}; see also the mean field treatment in
\cite{Dickman1999}. At lower temperature, $T_V$ indicates the spin
ordering (``vacuum order'') first reported in \cite{Melko2001a}.
Above $T_C(\rho)$ and small $\rho$, the system exists in a fluid
phase characterised by a homogeneous average local density of
monopoles and no long range charge order. As we increase the density
we reach a bifurcation point near $\rho=0.3$ where the crystallisation
transition splits in two: $T_S(\rho)$ marks the onset of SCO
with homogeneous local density, while the system phase separates at
$T_C(\rho)$ \cite{Note2}.

\begin{figure}[h]
\centerline{\includegraphics[angle=270,width=\columnwidth]{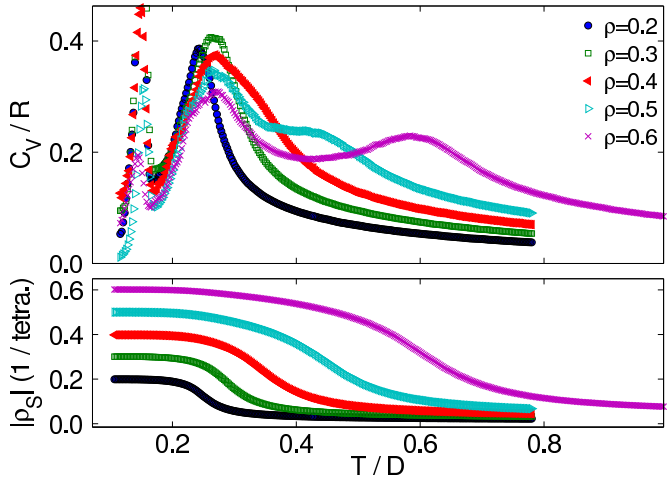}}
\caption{\emph{Upper panel.} Molar specific heat ($C_V$) as a function of
$T$ for $L=4$, at densities indicated in the legend. At very low
temperatures, a peak at $T_V$ indicates the ordering of the vacuum
even at high $\rho$. A broader peak at a higher temperature, $T_C$, signals
a transition in the charge degrees of freedom.  For $\rho > 0.3$ this 
transition splits into two: one at $T_C$ and a second one at $T_S$ . \emph{Lower panel.}
Staggered density $\rho_S(T)$ for the same densities.   For $\rho >
0.3$, the onset of SCO coincides with the high temperature
peak in $C_V$.} \label{fig3}
\end{figure}

Figure 4 has a startlingly similar counterpart in systems of simple
electric charges in a lattice, where $T_S$ was identified as a
second order N\'eel-like transition temperature, meeting at a
tricritical point with a first order dome where a ionic crystal
coexists with a low density disordered phase
\cite{Dickman1999,Dickman2000,Ciach2003}. A finite size scaling
analysis of our results confirms that $T_S$ is a second order
transition within the 3D-Ising universality class (see Sup.\ Info.\
for details).  Beyond these similarities, one fundamental difference
is that in our case the true degrees of freedom are spins, and the
presence of the charges --and their ordering-- are emergent
phenomena. 

A broader view is gained by comparison with other spin
systems. As in our case, the melting of the ionic crystal in the
uniformly-frustrated XY spin model in the triangular lattice
can take place in two stages .  However, this is a two-dimensional
system with a continuous symmetry and therefore one of these transitions
is of a different character (BKT) \cite{Korshunov}.  The resemblance
to our system is closer in the case of the Blume-Emery-Griffiths Model
used to describe He$_3$-He$_4$ mixtures \cite{chaikin2000principles},
and it extends even to the universality class expected for the tricritical
point.

The approximate character of the monopole picture implies that there
can be order not only in the internal structure of the monopoles
vacuum, but also within the magnetic charges themselves \cite{Note3}.
Indeed, we checked that different spin
configurations leading to \emph{the same} monopole cluster can have
different energies, and we were able to find a good candidate for
the spins ground state of the magnetic crystal, by exploring all
possible spin configurations within the conventional cubic unit cell
that would generate a perfect zincblende structure. Among the 48
configurations satisfying this condition, 16 had minimum energy,
differing from the next low lying energy level by about $6\%$. As
illustrated in the top inset to Fig.\ 1, the symmetry-connected
ground states are characterized by having zero magnetisation, with
the magnetic moments of each of the four up tetrahedra pointing
along the four possible $\langle 111 \rangle$ diagonals.

\begin{figure}[h]
\centerline{\includegraphics[angle=0,width=\columnwidth]{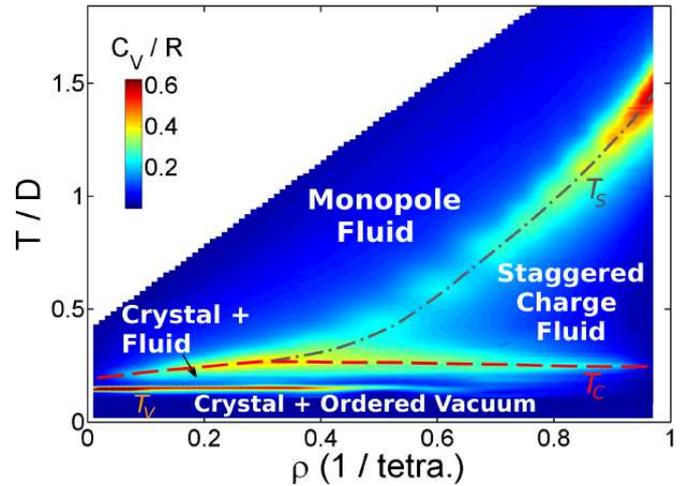}}
\caption{$T$ vs.\ $\rho$ phase diagram for the dipolar spin-ice. At
high temperatures, the system exists in a fluid phase of monopoles with
no long range order. As $T$ is lowered at low $\rho$ the fluid undergoes a
first order phase transition at $T_C(\rho)$, and crystallises into
the zincblende structure.  At  higher densities, the crystallisation
develops in two stages: the onset of staggered order at $T_S$ and phase
separation at $T_C$. A tricritical point at $\rho \approx 0.3$ and
$T/D \approx 0.26$ marks the meeting point of $T_C$ and $T_S$. 
Below $T_V$, the spin degrees of freedom of the
vacuum undergo an ordering transition. The diagram is overlaid on
top of an interpolated contour plot of $C_V$ for $L=4$ (see Fig.\
3). } \label{fig4}
\end{figure}

Our $\rho - T$ phase diagram in Fig.\ 4 can be related to that measured in
\DyO{} and \HoO{} under a field ${\bf H} // [111]$
\cite{Sakakibara2003,Higashinaka2004,Castelnovo2008,Krey2012}.  A
field in this direction induces a polarized state with 
monopoles in a crystalline zincblende structure and $\rho = 1$
\cite{Castelnovo2008}.  At low temperatures this phase is accessed
through a first order metamagnetic phase transition for fields of
the order of 1 tesla. This curve of metamagnetic transitions in the
$H - T$ phase diagram corresponds to the dome of first order phase
transitions in the $\rho - T$ phase diagram, $T_C$, in the same way
that the first order transition line in the pressure -- temperature
diagram for a vapour-liquid transition becomes a \emph{region} in
the temperature --density plane. The tricritical point in the $\rho
-T$ diagram  corresponds to the critical point of the first order
transition line in $H-T$ and the line of second order transitions
$T_S$ marking spontaneous symmetry breaking is replaced by a line of
crossovers in the $H-T$ plane, since $H$ is a symmetry breaking
field \cite{Sakakibara2003,Castelnovo2008,Krey2012}.

In summary, using the dipolar SI model we have been able to explore
the whole $\rho$ vs. $T$ phase diagram, observing charge-like
ordering in a purely magnetic system. Our model excludes the
possibility of double charges, a fact that limits its application to
all possible SI materials and temperatures. However, it has served
to explore a complex phase diagram, stressing in a unified view both
the monopolar nature of the excitations  and the
unavoidable need to take into account their spin nature at very low
temperatures (where the charge vacuum orders). The model also
provides a simple route to avoid the dynamical freezing while
retaining single spin-flip dynamics.

\begin{acknowledgments}
We thank R. Moessner, C. Castelnovo, T. S. Grigera, D. Cabra and G. Rossini
for helpful discussions, and the financial support of CONICET and
ANPCYT (Argentina), and the Royal Society (UK).
\end{acknowledgments}

\end{document}